\documentstyle[epsfig,twoside,fleqn,espcrc2]{article}

\def\ltsima{$\; \buildrel < \over \sim \;$}
\def\simlt{\lower.5ex\hbox{\ltsima}}
\def\gtsima{$\; \buildrel > \over \sim \;$}
\def\simgt{\lower.5ex\hbox{\gtsima}}

\newcommand{\AmS}{{\protect\the\textfont2
  A\kern-.1667em\lower.5ex\hbox{M}\kern-.125emS}}

\title{A Study of the Temperature gradient in Virgo/M87 with the MECS on board Beppo-SAX}

\author{F. D'Acri\address{CNR - Istituto di Fisica Cosmica e Tecnologie Relative \\ 
        Via Bassini 15, 20133 Milano, Italy}
        , 
        S. De Grandi\address{Max-Planck-Institut f$\ddot{\rm u}$r Extraterrestriche Physik\\
        D-85740 Garching, Germany}\\
        and S. Molendi$^{\rm a}$.\\}

\begin{document}

\begin{abstract}

Ground and in flight calibrations of the MECS experiment on board Beppo-SAX have
demonstrated that this is currently the best X-ray imaging experiment above 3 keV. The
MECS on-axis PSF has a half power radius of about 1 arcmin. Moreover due to a
fortunate combination of detector and mirror PSFs the total PSF depends only  weakly
on the energy. Finally the degradation of the PSF with off axis angle is negligible
within an off-axis angle of 10 arcminutes. Encouraged by these results we
developed techniques to analyze galaxy clusters observed with Beppo-SAX. In this proceeding
we quantify spectral distortions introduced by the energy dependent PSF when
performing spatially resolved spectroscopy of the core of the Virgo cluster.

\end{abstract}

\maketitle

\section{The MECS Point Spread Function}

The Medium Energy Concentrator Spectrometer (MECS; Boella et al. 1997/a) is one of the four 
narrow
field instruments on board the Beppo-SAX satellite (Boella et al. 1997/b).
The MECS operates in the energy band 1.3-10 keV with a field of view of $28'$
 radius.
The MECS consists of three units each composed of a grazing incidence Mirror
Unit (MU), and of a position sensitive Gas Scintillation Proportional Counter
(GSPC).\\
The Point Spread Function of the MECS (PSF$_{\rm MECS}$) is the convolution between the
MU PSF and the detector PSF.
The MU and detector PSFs are described, respectively, by a lorentzian function L(r) and
a gaussian function G(r).
Both the lorentzian and the gaussian functions are energy-dependent.
Typically the detector PSF improves with 
increasing energy, whereas the MU PSF improves with decreasing energy.
The detector PSF dominates the core of the PSF$_{\rm MECS}$ ($r \simlt 2'$) whereas the MU PSF 
dominates the wings of the PSF$_{\rm MECS}$ ($ \simgt 2'$).\\
The analytical expression for the on-axis PSF as given in Boella et al. 1997/a is:

$${\rm
PSF_{MECS}}(r,E)=\frac{1}{2\pi\Big[R(E)\sigma^{2}(E)+\frac{r^{2}_{l}(E)}
{2(m(E)-1)}\Big]}\cdot $$


\begin{equation}
\Bigg\{R(E)\exp\Bigg(-\frac{r^2}{2\sigma^2(E)}\Bigg)+\Bigg[1+\Bigg(\frac{r}
{r_l(E)}\Bigg)^2\Bigg]^{-m(E)}\Bigg\},
\end{equation}

where $R(E)$, $\sigma(E)$, $r_{l}(E)$ and $m(E)$ are algebraic functions of the energy E.\\
The integral of the PSF over the entire plane is normalized to unity:\\
$2\pi\int_{0}^{\infty}{\rm PSF_{\rm MECS}}(r,E)r\,dr\equiv1$.\\
We used eq. 1 to evaluate the $50\%$ and $80\%$ power radii
($r_{50}(E)$ and $r_{80}(E)$)  as a function of E:\\

\vskip -0.5truecm

$$2\pi\int_{0}^{r_{50}(E)}{\rm PSF_{\rm MECS}}(r,E)r\,dr=0.5,$$
\begin{equation}
2\pi\int_{0}^{r_{80}(E)}{\rm PSF_{\rm MECS}}(r,E)r\,dr=0.8\,.
\end{equation}

\begin{figure}[htb]
\vspace{9pt}
\epsfig{figure=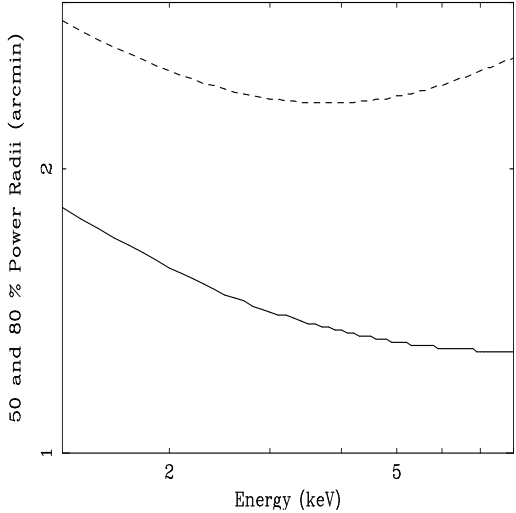, height=7.5cm, width=7.5cm, angle=-90}
\caption{ {\rm PSF}$_{\rm MECS}$ Power Radii vs. Energy. The solid line represents the 50 \% power
radius, r$_{50}$(E), the dashed line represents the 80 \% power radius, r$_{80}$(E) (see eq. 2).}
\end{figure}

As shown in fig. 1 $r_{50}(E)$ is always $<2'$ and decreases with increasing energy.
This is because at radii $<2'$ the PSF$_{\rm MECS}$ is dominated by the gaussian PSF of 
the detector, that improves with increasing energy.
The power radius $r_{80}(E)$ does not vary strongly with energy (see fig. 1)
because of the combined effect of the improvement of the detector PSF and degradation of the MU PSF
with increasing energy.

\begin{figure}[htb]
\vspace{9pt}
\epsfig{figure=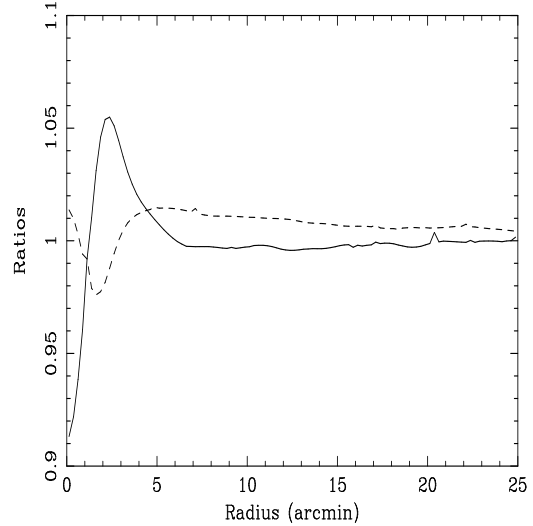, height=7.5cm, width=7.5cm, angle=-90}
\caption{ Ratios between convolved profiles vs. radius r. Solid line: Ratio R$_{1}$(r) between
the convolved profile $\tilde{I}$(r,3 keV) and the convolved profile $\tilde{I}$(r,6 keV). 
Dashed line: Ratio R$_{2}$(r) between the convolved profile $\tilde{I}$(r,9 keV) and the convolved 
profile $\tilde{I}$(r,6 keV)}
\end{figure}

\begin{figure}[htb]
\vspace{9pt}
\epsfig{figure=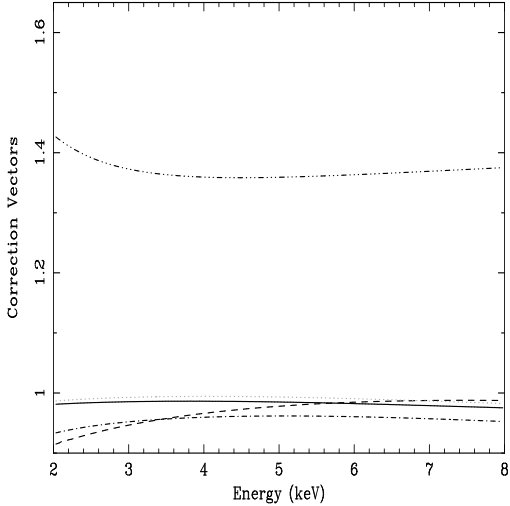, height=7.5cm, width=7.5cm, angle=-90}
\caption{ Correction vectors for the Virgo cluster observed with the MECS. Dot-dot-dot-dashed line:
correction vector V$_{1}$(E$_{i}$) for the region $0'-2'$. Dashed line: correction vector
$V_{2}$(E$_{i}$) for the region $2'-4'$. Dash-dotted line: correction vector $V_{3}$(E$_{i}$) for 
the region $4'-6'$. Dotted line:  correction vector $V_{4}$(E$_{i}$) for the region $6'-8'$.
Solid line:  correction vector $V_{5}$(E$_{i}$) for the region $8'-10'$.  }
\end{figure}

\section{Convolution of the Source Radial Profile with the PSF$_{\rm MECS}$}

A proper analysis of extended sources, like clusters of galaxies, requires
that the blurring introduced by the limited spatial resolution of the observing
instruments be correctly taken into account.
In practice, this is done by evaluating the convolution of the source surface
brightness profile, $I$, with the PSF$_{\rm MECS}$. We have approximated $I$ with the profile
of the Virgo cluster as observed with the PSPC instrument on board the ROSAT satellite,
because of the considerably better spatial resolution of this instrument respect to the
MECS.\\
The ROSAT Virgo cluster profile, $I_{\rm ROSAT}(r_{0})$, convolved with the PSF$_{\rm MECS}$
at any point P($r$) in polar coordinates ($r_{0},\varphi$) is:\\

\begin{figure}[htb]
\vspace{9pt}
\epsfig{figure=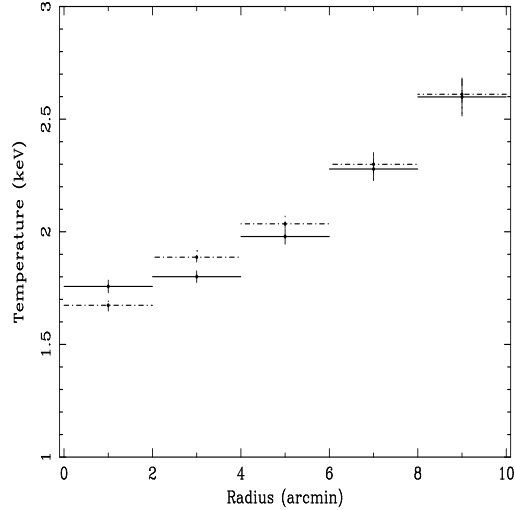, height=7.5cm, width=7.5cm, angle=-90}
\caption{ Virgo temperature profiles. The solid line is the temperature profile $T_{obs}(r)$ 
obtained from the observed spectrum, $S_{obs}$, whereas the dash-dotted line is the temperature 
profile $T_{corr}(r)$ obtained from the corrected spectrum, $S_{corr}$.}
\end{figure}

\noindent

$\tilde{I}(r,E)=\int_{0}^{\infty}dr_0\cdot$

\begin{equation}
\int_{0}^{2\pi}d\varphi\,I_{\rm ROSAT}(r_0)\,
{\rm PSF_{\rm MECS}}(r,r_{0},\varphi,E)\,r_0\,.
\end{equation}

We computed the convolved profiles $\tilde{I}(r,3{\rm keV})$, 
$\tilde{I}(r,6{\rm keV})$ and $\tilde{I}(r,9{\rm keV})$.\\
In order to estimate the spectral distortions introduced by the energy-dependent
PSF$_{\rm MECS}$ we calculated the ratios $R_{1}(r)=\frac{\tilde{I}(r,3{\rm keV})}
{\tilde{I}(r,6{\rm keV})}$ and $R_{2}(r)=\frac{\tilde{I}(r,9{\rm keV})}
{\tilde{I}(r,6{\rm keV})}$.\\
As shown in fig. 2, $R_{1}$ and $R_{2}$ are contained 
within 0.09 of unity for any radius and within 0.06 for radii $>1'$.
The obvious implication is that spectral distortions introduced by the PSF$_{\rm MECS}$
are quite modest.

\section{Spectra Correction Method}

The spectra obtained from the MECS observations of clusters of galaxies are affected by
the blurring effects of the PSF$_{\rm MECS}$. To correct these effects we
computed correction vectors $V(E_{i})$ where i=1,..,256 are the energy channels of the MECS. 
These vectors 
$V(E_{i})$ are quantities that, multiplied by the spectrum of the cluster
observed by the MECS, $S_{obs}(E_{i})$, give us the corrected spectrum of the cluster,
$S_{corr}(E_{i})$:


\begin{equation}
S_{corr}(E_{i})=S_{obs}(E_{i})V(E_{i}).
\end{equation}

We computed correction vectors for annular regions centered on the emission peak of the
Virgo cluster. We considered 5 annular regions with inner and outer radii of $0'-2'$,
$2'-4'$, $4'-6'$, $6'-8'$ and $8'-10'$ respectively. The correction
vector for the $j_{th}$ region spectrum, $V_{j}(E_{i})$, is:

\begin{equation}
V_{j}(E_{i})=\frac{\int_{2(j-1)}^{2j}2\pi\,r
I_{\rm ROSAT}(r,E_{i})dr}
{\int_{2(j-1)}^{2j}2\pi\,r\,\tilde{I}(r,E_{i})dr}
\end{equation}

where $\tilde{I}(r,E)$ is defined by eq. (3).\\
All the correction vectors evaluated for the considered regions are shown in fig. 3.\\
We note that: 1) $V_{1}(E_{i})$ has a mean value of $\sim 1.4$, while all other
correction vectors are contained between 0.9 and 1.0. This is because the number of
photons revealed is lower than the number of the photons emitted by the source, while the viceversa
is true for the others four regions.
2) All correction vectors show small variations with the energy and therefore the spectral
distortions are modest.

\section{Temperature Gradients Measures in Virgo/M87}

The observed, $S_{obs}$, and corrected, $S_{corr}$, spectra obtained from the annular
regions described in the previous section have been used to measure temperature
profiles.
This has been done by fitting each spectrum in the energy range 1.4-5 keV with thermal
emission model (MEKAL in XSPEC ver. 9.01). In fig. 4 we show the temperature profiles
$T_{obs}(r)$ and $T_{corr}(r)$ obtained from the observed, $S_{obs}$, and the corrected, 
$S_{corr}$, spectra.
The difference in value  between $T_{obs}(r)$ and $T_{corr}(r)$ are negligible.
This demonstrates that the effects introduced by the PSF$_{\rm MECS}$ are very small.

We compared our results with those of the ROSAT and ASCA satellites.
To make this possible we fitted the spectra with the same models used by Nulsen and 
B$\ddot{\rm o}$hringer (1995) in the analysis of the ROSAT PSPC data and by Matsumoto 
et al. (1996) in the analysis of the ASCA  GIS data:  
Nulsen and B$\ddot{\rm o}$hringer used a Raymond-Smith model (Raymond \& Smith 1977) in the energy
band 0.5-2.4 keV,
while Matsumoto et al. used a thermal bremsstrahlung model plus gaussian line in the energy band
3-10 keV.
We recall that for $r<6'$ Nulsen and B$\ddot{\rm o}$hringer found temperatures in the energy range 
1.1-2.3 keV, while Matsumoto et al. found temperatures in the energy range 2.-2.3 keV. 
Fitting the $S_{corr}$ spectra with a Raymond-Smith model in the energy range 1.4-2.4 keV we found, 
basically, the same temperatures as in the ROSAT analysis.
Using a bremsstrahlung model plus a gaussian line in the energy range 3-10 keV we found temperatures 
consistent with those found from the ASCA GIS analysis (see their table 1).   
Details about these comparisons are in D'Acri et al. (1998).


\end{document}